\documentclass[twocolumn,pra,superscriptaddress]{revtex4}
\usepackage{amsmath}
\usepackage{graphicx}
\usepackage{color}
\usepackage{graphicx}
\usepackage{amssymb}
\usepackage{bm}
\usepackage{subfigure}
\newcommand{\hide}[1]{}

\newcommand{\eq}[1]{Eq.\,(\ref{#1})}

\newcommand{\fig}[1]{Fig.\,\ref{#1}}

\newcommand{\ket}[1]{\ensuremath{\left| #1 \right\rangle}}

\begin{document}

\title{Studies on Negative Refractive Index without Absorption}
\author{R. M. Rajapakse}
\affiliation{Department of Physics, University of Connecticut,
Storrs, CT 06269}
\author{E. Kuznetsova}
\affiliation{Department of Physics, University of Connecticut,
Storrs, CT 06269}\affiliation{ITAMP, Harvard-Smithsonian Center
for Astrophysics, Cambridge, MA 02138}
\author{S. F. Yelin}
\affiliation{Department of Physics, University of Connecticut,
Storrs, CT 06269} \affiliation{ITAMP, Harvard-Smithsonian Center
for Astrophysics, Cambridge, MA 02138}

\date{\today}

\begin{abstract}
Which systems are ideal to obtain negative refraction with no absorption? Electromagnetically induced transparency (EIT) is a method to suppress absorption and make a material transparent to a field of a given frequency. Such a system has been discussed in \cite{Jurgen}; however the main limitations for negative refraction introduced are the necessity of resonant electric and magnetic dipole transitions, and the necessity of very dense media.   We suggest using frequency translators in  a composite system that would provide negative refraction for a range of optical frequencies while attempting to overcome the limitations discussed above.  In the process of using frequency translators, we also find composite systems that can be used for refractive index enhancement.
\end{abstract}
\pacs{03.65.Ud, 03.67.Mn, 42.50.-p, 42.50.Dv}
\maketitle
\section{Introduction}
Negative refraction has been a topic of interest since the 1960s \cite{Veselago} and has become an active area of research in modern photonics.  The development of a perfect lens where the imaging resolution is not limited by wavelength is an application of much interest \cite{Veselago}.   After the initial proposal for realization in meta-materials \cite{Pendrytwo}and experimental demonstration \cite{Shelby} for RF radiation, much progress has been made towards negative refraction for shorter and shorter wavelengths \cite{Shalaev,Soukoulis,Shelby,Yen,Linden,Parimi,Berrier}. The approaches include split-ring resonator meta-materials \cite{Shelby, Yen, Linden} and photonic crystals \cite{Parimi, Berrier} .  Much development has been made in this field to date \cite{Shelby,Yen,Linden} but creating negative refraction with minimal absorption has been a challenge.

Although it was first suggested that negative refractive index required media with both negative permittivity and permeability ($\epsilon, \mu < 0$) in the frequency range of interest \cite{Veselago}, when the optical regime is considered it is difficult to realize negative permeability without considerable loss.  The transition magnetic dipole moments ($\mu_a$) are smaller than the transition electric dipole moments $d_a$ by a factor of the order of the fine structure constant $\alpha$. As a result, the magnetic susceptibilities are much smaller than electric susceptibilities.  This condition can be alleviated by using a material such as a chiral medium where the electric and magnetic properties are cross-coupled \cite{Pendry}.

Enhancement of refractive index has also been a topic of interest \cite{YavuzSimmons}, and experimental results have shown that such enhancement can be achieved with no absorption, using electromagnetically induced transparency (EIT) \cite{YavuzSimmons}.  The use of EIT to suppress absorption  has been suggested by refs \cite{Jurgen,Jurgen2}, and several systems which can be used to achieve electromagnetically induced negative refraction without absorption are discussed in this paper.  We propose to extend the work done by \cite{Jurgen2}.

The scheme used in \cite{Jurgen,Jurgen2} is based on coupling the electric and magnetic components of the probe field to different transitions in the atom, one electric and one magnetic dipole transition. Thus, the  main limitations of introducing atomic systems for negative refraction are the necessity of resonant electric and magnetic dipole transitions, and the need of very dense media. This latter condition stems mostly from the fact that all elements in the scheme involving magnetic dipole coupling are by nature very weak. Many atomic, bound exciton or polar molecular schemes have few resonant levels for electric and magnetic dipole transitions, especially in the range of optical frequencies because magnetic quantum transitions require the same electronic configuration, i.e the same $n$ and $l$ quantum numbers, which can for optical frequencies only be reached for systems with high relativistic effects, i.e., involving very heavy atoms. 

 In order to obtain sufficiently large magnetic dipole transition frequencies for example, to obtain a visible dipole transition, it is also necessary to use heavy atoms.   Even so, as there are few resonant levels  we attempt to implement a coherent frequency translator in order that electric and magnetic dipole coupling can occur between two levels with different optical frequencies while leaving the phase intact.  We then revise the system which includes the frequency translator into a different level structure which decreases the absorption and has negative refraction over a larger range of detuning.

\section{Negative refraction via electromagnetically induced chirality}

We begin by revisiting some of the concepts of \cite{Jurgen}.  The refractive index of a medium can be expressed by the following equation:
\begin{equation}
n=\sqrt{\epsilon \mu -\frac{(\xi_{EH} +\xi_{HE})^2}{4}}-\frac{i}{2}(\xi_{EH}-\xi_{HE})
\label{originalequation}
\end{equation}
where the values of $\epsilon$, $\mu$, $\xi_{HE}$ and $\xi_{EH}$ can be found from \eq{eqPM} \cite{ODell}.

\begin{figure}[ht]
\subfigure[ ]{
\includegraphics[trim =50mm 10mm 20mm 20mm, clip,width=0.2\textwidth]{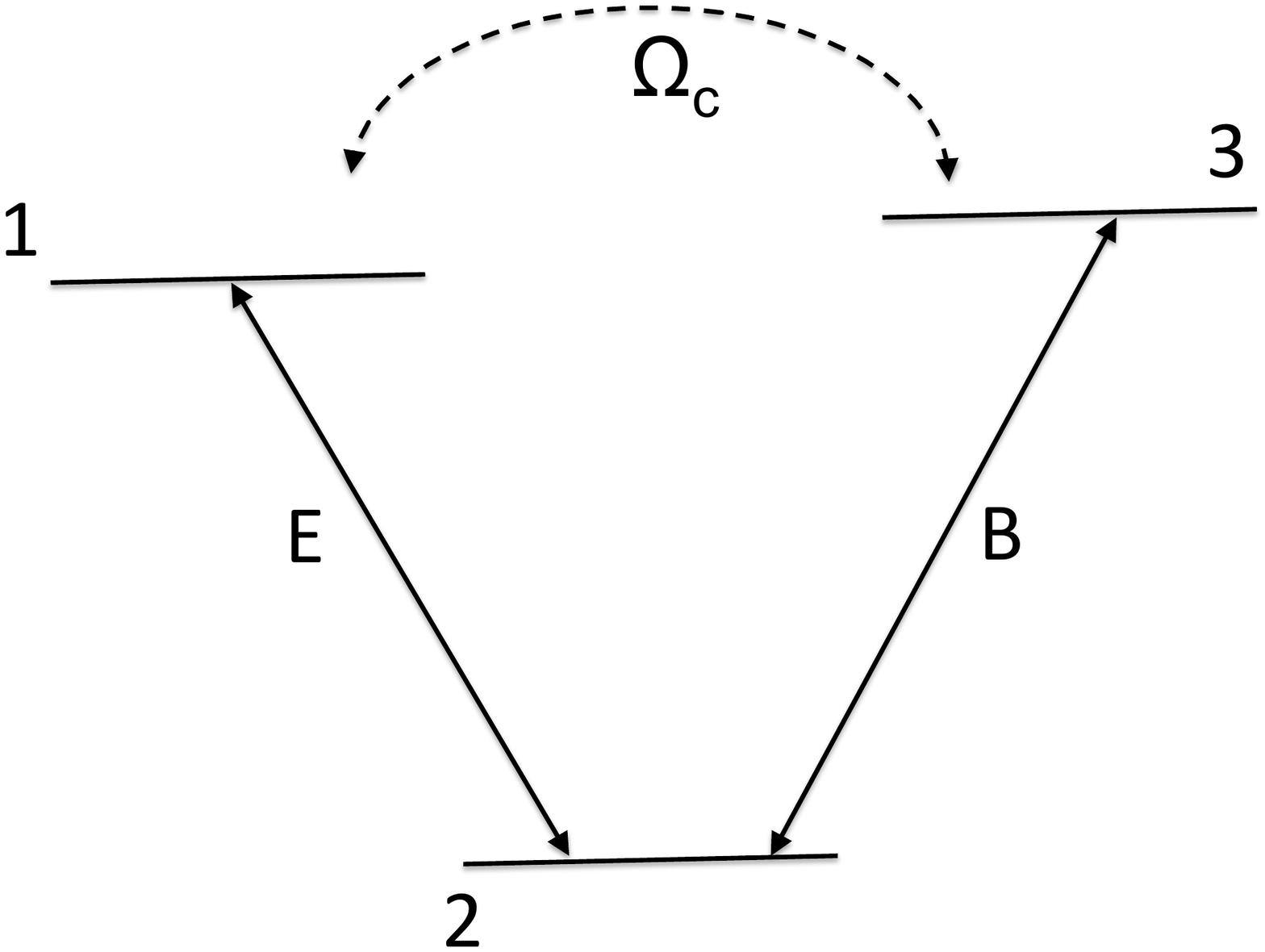}
\label{fig:threelevel}
}
\subfigure[ ]{
\includegraphics[trim = 40mm 20mm 35mm 30mm, clip,width=0.25\textwidth]{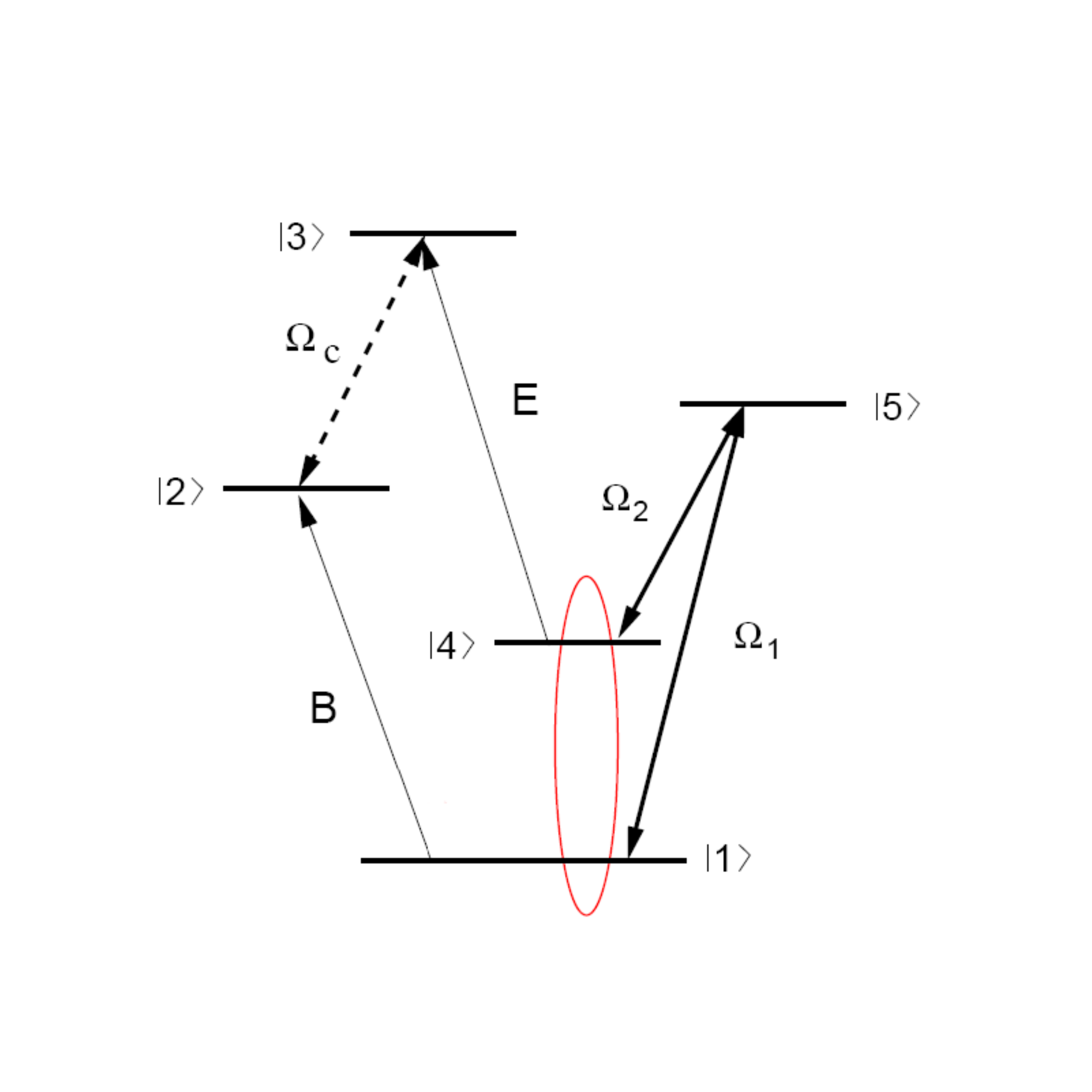}
\label{fig:fivelevel}
}
\label{fig:threefivelevel}
\caption{a) Level scheme for obtaining electromagnetically induced chirality. b) Modified level scheme for obtaining electromagnetically induced chirality \cite{Jurgen}. The dark state
created by the two-photon resonant Raman coupling of levels $\ket{1}$
and $\ket{4}$  takes over the role of level$\ket{1}$ in (a). Both level
schemes have E and  B as the electric and magnetic components of the
probe field, which can experience negative refraction.}
\end{figure}

A simple three level system (see \fig{fig:threelevel}) can be used to introduce electromagnetically induced chiral negative refraction.  Here the electric and magnetic components of the probe field couple state $\ket{1}$ to $\ket{2}$ by an electric dipole transition and state $\ket{3}$ to state $\ket{2}$ by a magnetic dipole transition.  a strong coherent field couples $\ket{1}$ and $\ket{3}$ with Rabi frequency $\Omega_c$. Level $\ket{3}$ can be considered metastable with a decay rate $\gamma_3 \sim (1/137^2)$ $\gamma_1$. 


In realistic media this setup can be modified using three criteria (i) $\Omega_c$ must be an ac field so its phase can be adjusted to induce negative refraction. (ii) There must be high contrast EIT for the probe field in order to suppress absorption and (iii) the energy level structure must be appropriate for media of interest as the parities should be conducive to introducing the appropriate electric and magnetic dipole coupling.  \fig{fig:fivelevel} is an example that meets the above criteria.  This scheme employs strong coherent Raman coupling by two coherent fields with complex Rabi fields $\Omega_1$ and $\Omega_2$ and carrier frequencies of $\omega_1$ and $\omega_2$, creating a superposed dark state of the two levels $\ket{1}$ and $\ket{4}$. The dark state acts as the ground state and the modified scheme is effectively the same as with \fig{fig:threelevel} (which is taken from \cite{Jurgen}) but with more degrees of control.  


\section{Limitations of current approach}
For the scheme described in \fig{fig:fivelevel} we use a medium which has the cross couplings given in \eq{PMcoupling} \cite{ODell} to obtain negative refraction, with
\begin{eqnarray}\label{PMcoupling}
\mathbf{P}&=&\epsilon_0 \chi_e \mathbf{E} +\frac{ \xi_{EH}}{c} \mathbf{B}\nonumber\\
\mathbf{M}&=& \frac{\xi_{HE}}{\mu_0 c} \mathbf{E} + \chi_{m}\mathbf{B}
\label{eqPM}
\end{eqnarray}

The above equations require that there should be a coupling between the electric and magnetic fields in the medium, which have to be components of the same input field.

The requirement of a single input field imposes some limitations:  the levels being coupled by the electric field and the levels coupled by the magnetic field should then have the same energy difference and opposing parities.  Level selection in such a situation is complicated; magnetic dipole transitions leave electronic configurations intact and it is difficult to find a material in which this condition is satisfied.  Heavy atoms have relativistic effects which satisfy the conditions of changing the electronic configuration. Rare-earth materials such as Dysprosium vapor, Scandium and Neodymium, therefore,  have a level structure that is conducive to obtaining negative refraction \footnote{ A change in principal quantum number n leads to much more forbidden magnetic dipole transitions}.  

The two problems of a difficult level selection and the requirement of high densities, which are of the order of $\rm{10}^{17} \rm{cm}^3$ for typical parameters  \cite{Jurgen} have to be solved to extend the types of materials which can be used to obtain negative refraction.  We consider composite systems which involve frequency conversion to overcome the requirement of having the electric and magnetic field have the same frequency.


\section{Composite systems for wider parameter regions}

To overcome the difficulty of finding a system where an allowed electric dipole transition is the exact same frequency as an allowed magnetic dipole transition, we attempt to expand our scheme in a way that allows dipole coupling of electric and magnetic components of different fields, but yet retains characteristics, particularly phase information of the original single field entering the media. Conceptually, the index of refraction indicates the response of a field to the initial atomic response, or the changes in phase velocity of the field.  As the field has an effect on the phases of both polarization and magnetization, the interplay of the electric and magnetic component of the field has to taken into account.  In order to achieve the result described above it is necessary to use a single initial field, divide it so that the phase is maintained, and later recombine the fields coherently and observe the response. 
\subsection{Proof of principle}


We require the one-to-one conversion of one field to another down to the single photon level model, while keeping the phase and coherence unchanged. The calculation below illustrates that adding  frequency conversion does not fundamentally change the dynamics of a single photon from the five level process described in \cite{Jurgen}.  $\rm{H}_{NI}$ is the Hamiltonian for  \fig{fig:fivelevel}, which describes the dynamics of a single photon in a five level medium. 

$g_1^{*} a^{\dag} \sigma_{34}$ and  $g_2^{*} a^{\dag} \sigma_{21}$ describe the coupling of the probe field to the two probe transitions.  $g_1^{*} a^{\dag} \sigma_{34}$ is an electric dipole transition while $g_2^{*} a^{\dag} \sigma_{21}$ is a magnetic dipole transition.
We obtain the single-photon Hamiltonian for \fig{fig:fivelevel}, which gives us:
\begin{eqnarray}
H_{NI} = \sum_{i=1}^{5} \Delta_i\sigma_{ii} + \delta a^{\dag}a + g_1^{*} a^{\dag}  \sigma_{34}+g_2^{*} a \sigma_{21} + \nonumber\\ 
\Omega\sigma_{23}+ \Omega_1\sigma_{45} 
+ \Omega_2\sigma_{15} + \rm{h.c.}
\label{HNI}
\end{eqnarray}
From the above, we obtain the following dynamics: 
\begin{equation}
 \partial a = \imath \delta a + \imath g_1^* \sigma_{34} + \imath g_2^*\sigma_{21}.
 \label{partial1}
 \end{equation}
The quantiy $\delta$ is related to the dispersion mismatch between pumps and sidebands \cite{McGuiness}.

One idea to create a division of a single field while maintaining the phase is to add a four wave mixing process in order to ensure that there is coupling between two fields that enter the four wave mixing medium and the fields of different frequency that exit it \cite{Johnsson2, Johnsson}.  We use four wave mixing in a composite system, and discuss how it affects the behavior of a single photon.

 For the composite system with frequency conversion, our Hamiltonian changes to
 \begin{eqnarray}
 H &=& H_{mix}+H_{NI} \nonumber\\
 H_{mix} &=& \delta_1a_1^{\dag}a_1 + \delta_2a_2^{\dag}a_2+\kappa a_1^{\dag}a_2 + \kappa^* a_2^{\dag}a_1 \nonumber\\
 H_{NI} &=&  \sum_{i=1}^{5} \Delta_i\sigma_{ii} + g_1^{*} a_1^{\dag}  \sigma_{34}+g_2^{*} a_2^{\dag} \sigma_{21} + \Omega\sigma_{23}+\nonumber \\&& \Omega_1\sigma_{45} + \Omega_2\sigma_{15} + \rm{h.c.}
 \end{eqnarray}
 By eliminating  $a_2$ adiabatically, this gives us
\begin{equation}
 \partial a_1 = \imath (\delta_1-\frac{|\kappa|^2}{\delta_2}) a_1 + \imath g_1^* \sigma_{34} - \imath \frac{\kappa}{\delta_2} g_2^*\sigma_{21}
 \label{partial2}
\end{equation}
where $\kappa$ is related to the nonlinearity \cite{McGuiness}.
From the similarity of \eq{partial1} and \eq{partial2} above, we can see that the addition of a frequency conversion Hamiltonian still gives us an outcome which is analogous to that of a single photon with no frequency conversion.  We use four wave mixing for the physical implementation of this frequency conversion.

\subsection{Four wave mixing}
Four wave mixing involves four electromagnetic fields interacting in a medium.  To keep the relationship between the electric and magnetic components of the field intact in a negative refractive media, we use an initial weak single field to create two fields of different frequencies that coherently couple both an electric dipole transition and a magnetic dipole transition in the medium we use.  This will ensure that two different fields are created that will carry the same phase information, and these fields will create separate electric and magnetic dipole transitions in the negative refractive part of the system corresponding to \fig{fig:fivelevel}, with a Hamiltonian of similar form as \eq{HNI}.

\begin{figure}[h]
\begin{center}
\leavevmode {\includegraphics[scale=0.35]{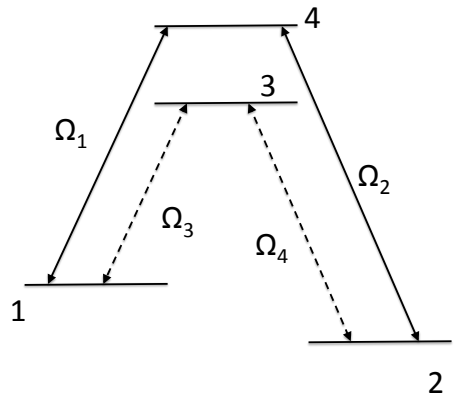}}
\end{center}
\caption{A four wave mixing setup to obtain an additional field.  One exiting field creates an electric dipole transition, and the other a magnetic dipole transition in \fig{fig:fivelevel}}\label{fourwave1}
\end{figure}

Our suggested four level mixing medium consists of a double $\Lambda$ system. Two control fields  in the four level system, $\Omega_1$ and $\Omega_2$ are the driving fields which keep the population in the mixed ground state, and the third anti-Stokes field is a field  $\Omega_3$ which couples levels $\ket{1}$ and $\ket{4}$ in \fig{fourwave1}.  

Mixing of these three fields creates a Stokes fourth field, $\Omega_4$ which will then couple two magnetic dipole allowed levels, with a frequency $\omega_4= \omega_1 + \omega_2 - \omega_3$. We specify that the two weak fields couple to a single upper level in order to minimize gain.

For the output, we consider two input fields (the original and the generated) that pass through another four wave mixing setup. 

\begin{figure}[h]
\begin{center}
  \subfigure[ ]{
 {\includegraphics[trim = 20mm 60mm 1mm 60mm, clip,scale=0.35]{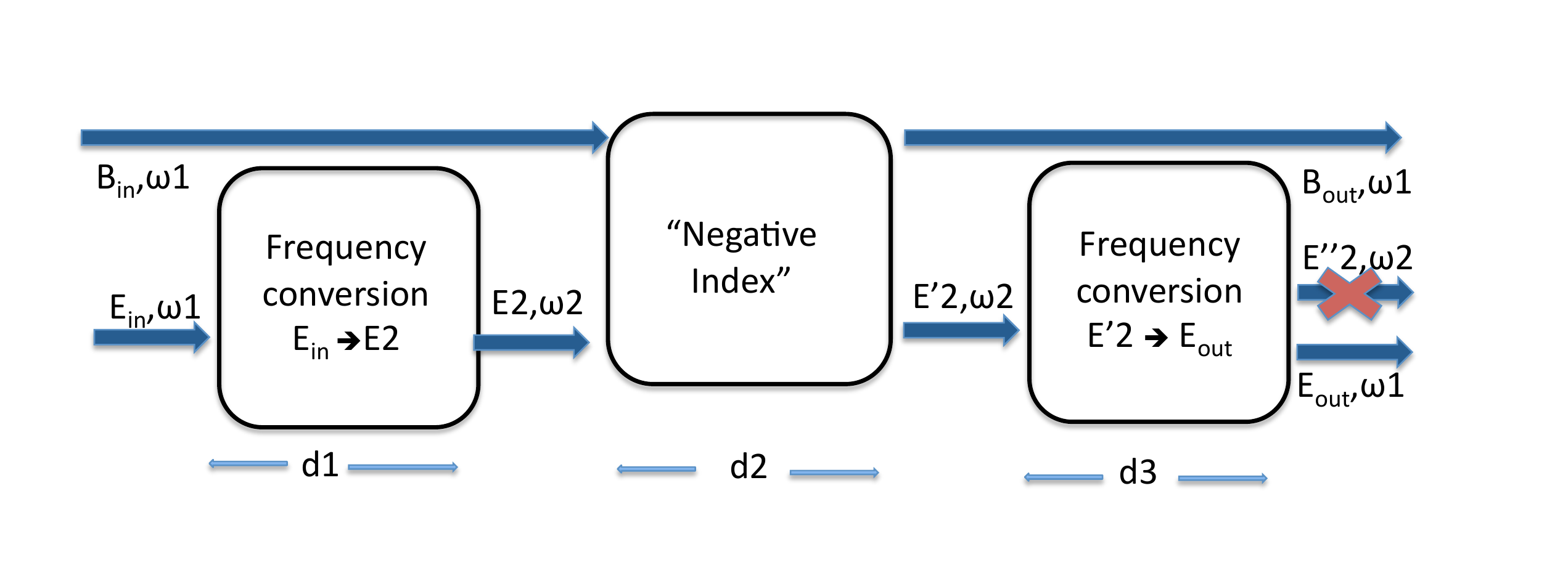}}
 \label{fourwavemix3}
 }\\
 \subfigure[]{
  {\includegraphics[trim =20mm 60mm 1mm 60mm, clip,width=0.5 \textwidth]{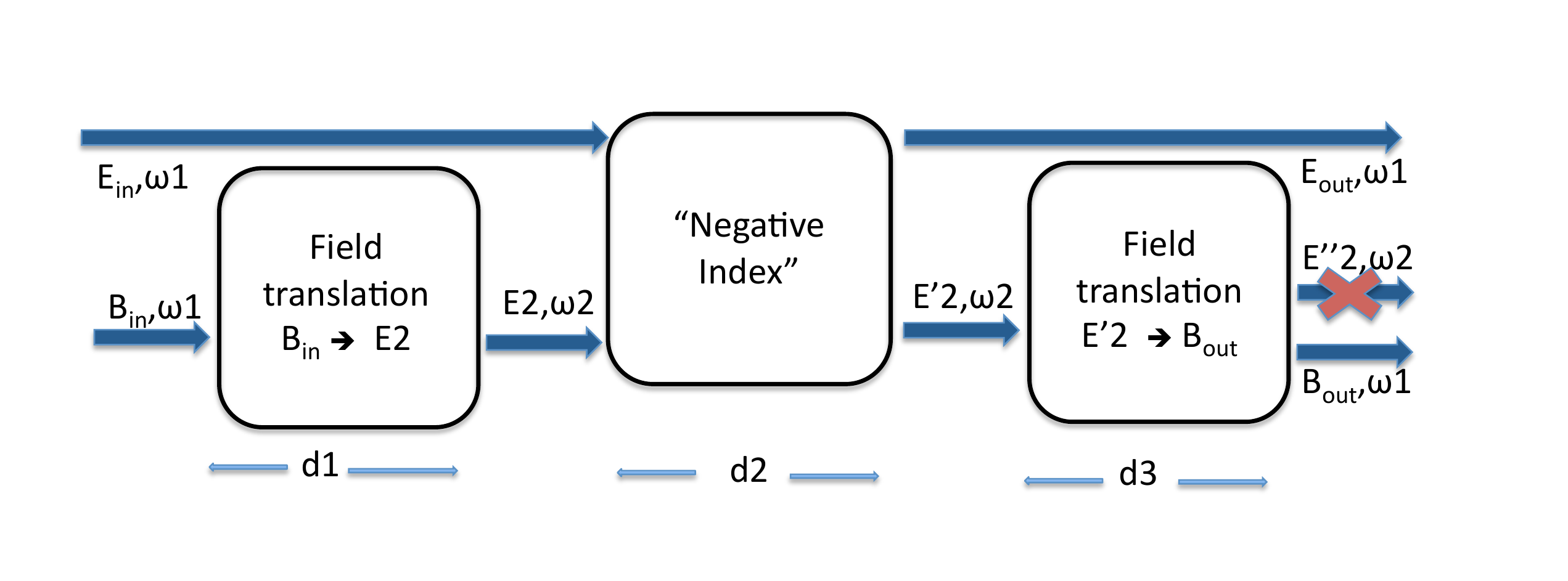}}
  \label{Boxesfor545}
  }
\end{center}
\caption{ \protect Two composite setups. \fig{fourwavemix3} translates the frequency of the electric field, while \fig{Boxesfor545} has translation between electric and magnetic fields. }

\end{figure}

All the fields must maintain a constant phase relative to the others:
\begin{equation}
\vec{k_1} + \vec{k_2}=\vec{k_3}+\vec{k_4}
\label{kvec}
\end{equation}
where any change in relative phase corresponds to a change in the frequency of the outgoing fields and is therefore not conducive to absorption-free four wave mixing.
\fig{fourwavemix3} shows the setup we suggest.

In materials of positive refractive index the Poynting vector is in the same direction as $\vec{k}$ but in negative index materials the Poynting vector moves in the opposite direction to $\vec{k}$ \cite{popov1}.   
Additional error occuring in four wave mixing because of phase noise is discussed in detail in \cite{LukinFourwave}. Atomic noise is contributed by the relaxation rates of the ground state coherence, and an additional noise contribution occurs from the absorption of the driving field.  Minimizing absorption within the four-wave mixing setup  can be achieved by adding a detuning of one of the driving fields from resonance and choosing the driving fields within certain parameters as calculated in \cite{LukinFourwave}.

\section{Simulation: effective refractive index}

The "effective refractive index" of a mixed system depends on the effective electric and magnetic susceptibilities of the complete system. In order to derive the effective refractive index of the material, we have to obtain the effective permittivity, the effective permeability and the effective cross-couplings of the combined system.   The effective susceptibilities given below are calculated assuming a composite system as in \fig{fourwavemix3} and \fig{Boxesfor545}, where all the transitions take place in a single combined system of thickness $d$, where $d=d_1+d_2+d_3$. We calculate the propagation matrices for the fields $E_{in}$ and $B_{in}$ in \fig{fourwavemix3} and find the equivalent matrix that transforms the input field to the output field. Detailed derivations are given in Appendix A. 

The effective susceptibilities of the composite system are given by
\begin{eqnarray}
\chi_{ee-eff}&=&\frac{c}{\imath \omega_1(d_1 + d_2 + 
  d_3)} (C_2 C_5 d_1 d_3 \chi_{21}\chi_{12}(1+\nonumber \\ &&d_2 F_4\chi_{ee}^{})-1)\nonumber\\
\chi_{eb-eff}&=&\frac{c}{\imath \omega_1(d_1 + d_2 + d_3)}d_1d_2 F_4 C_5 \chi_{eb}^{}\chi_{12}\nonumber\\
   \chi_{bb-eff}&=&\frac{c}{\imath \omega_1(d_1 + d_2 + d_3)}(d_2 F_3^{} \chi_{bb}^{})\nonumber\\
\chi_{be-eff}&=&\frac{c}{\imath \omega_1(d_1 + d_2 + d_3)}d_1 d_2 C_2^{} F_3^{}  \chi_{be}^{}\chi_{21}^{}
 \end{eqnarray}
 
 Here $d_1$, $d_2$ and $d_3$ are the thicknesses of the materials in the composite system, $C_2=F_4= \frac{\imath \omega_2}{c}$ and $C_5=F_3= \frac{\imath \omega_1}{c}$.
 
We then use the effective susceptibilities in \eq{originalequation} to obtain the refractive index.

\subsection{Results}

\begin{figure}[ht!]
     \begin{center}
        \subfigure[ ]{%
            \label{fig:NN15X10^16chieeeffJo}
            \includegraphics[width=0.2\textwidth]{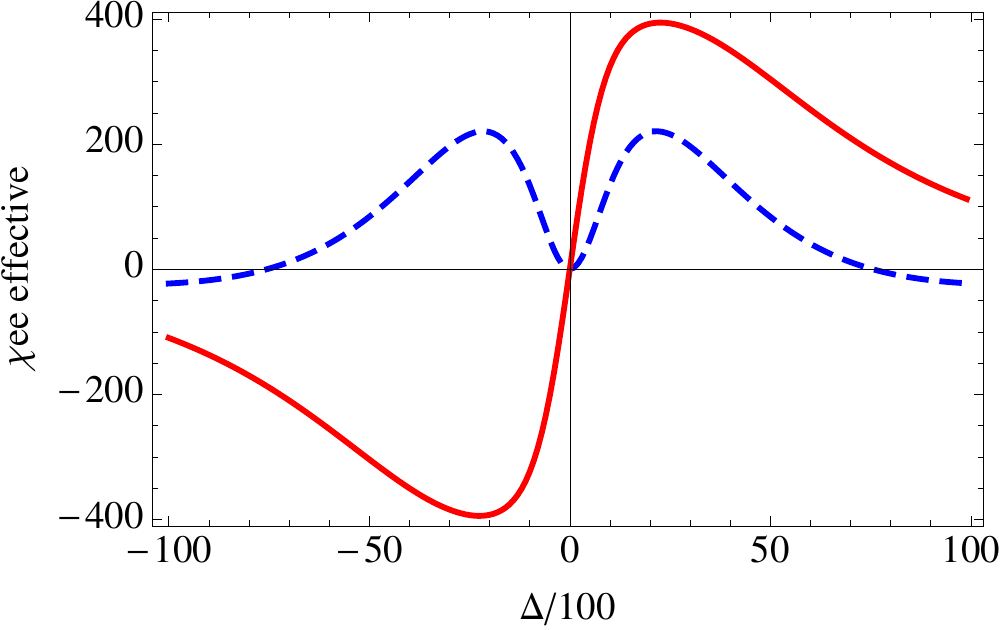}
        }%
        \subfigure[ ]{%
           \label{fig:NN15X10^16chiebeffJo}
           \includegraphics[width=0.2\textwidth]{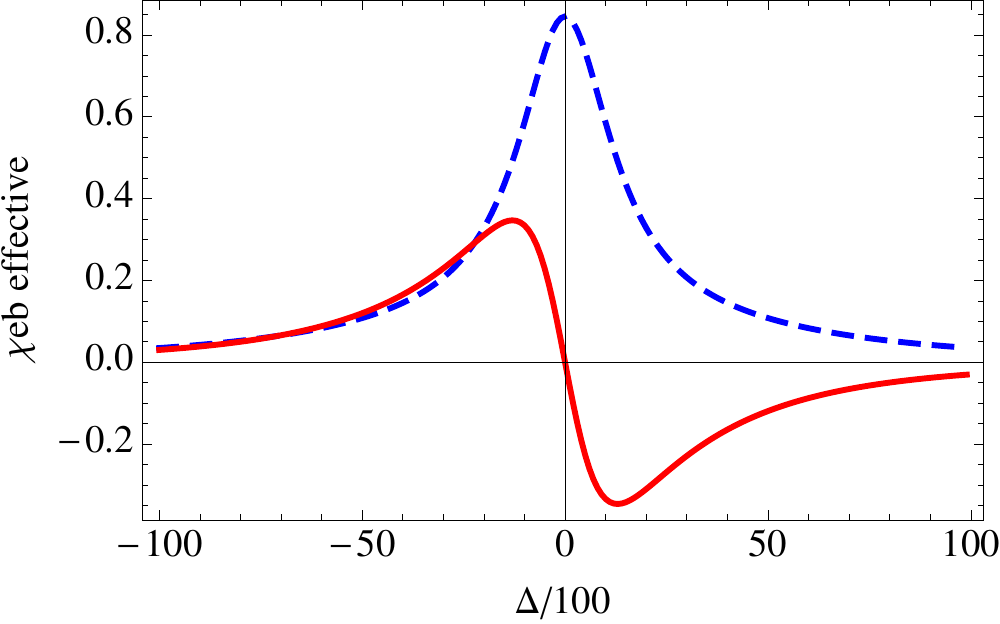}
        }\\ 
        \subfigure[ ]{%
            \label{fig:NN15X10^16chibeeffJo}
            \includegraphics[width=0.2\textwidth]{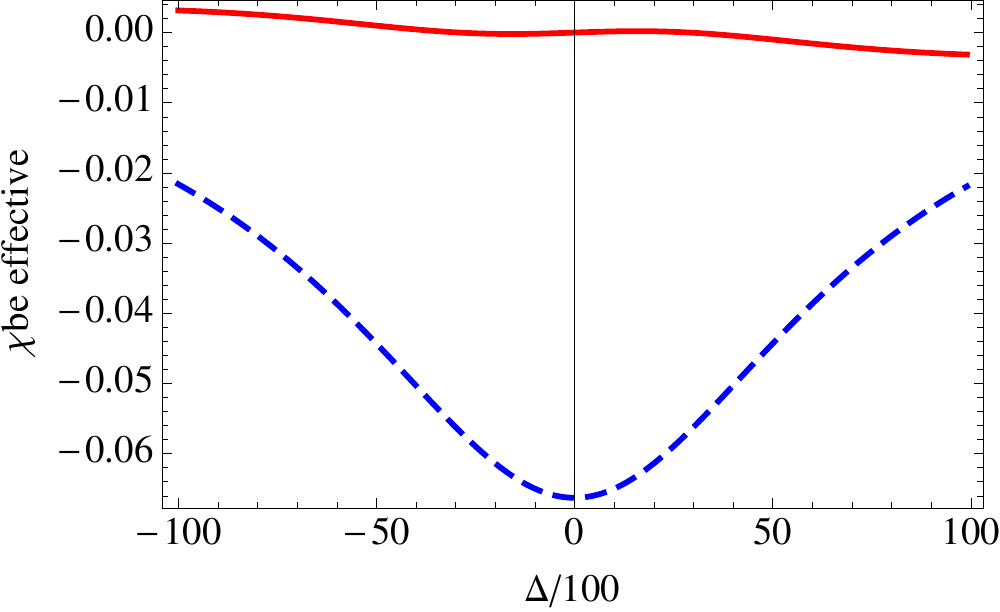}
        }%
        \subfigure[ ]{%
            \label{fig:NN15X10^16chibbeffJo}
            \includegraphics[width=0.2\textwidth]{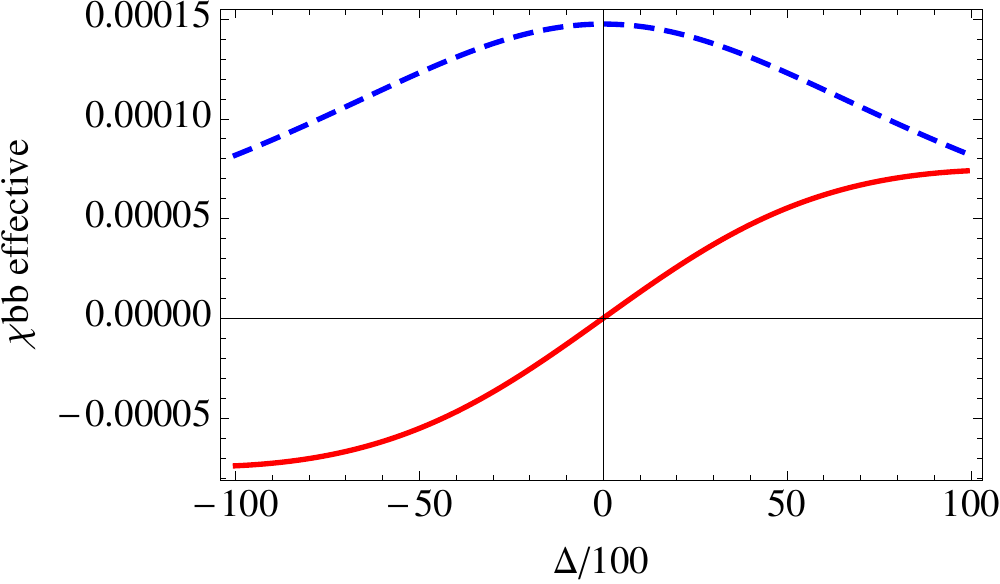}
        }
    \end{center}
    \caption{%
        (a)-(d) Real(red) and imaginary(blue-dashed) parts of the effective susceptibilities for the composite system with frequency translation of the electric field. Density of "negative refractive" system is $\rm{9}\times\rm{10}^{15}\rm{cm}^{-3}$, $d_1=d_3= 8 \times 10^{-2}\rm{cm}$, $d_2=8\rm{cm}$.
      } %
   \label{fig:454Susceptibilities4WM}
\end{figure}

\begin{figure}[ht!]
     \begin{center}
        \subfigure[]{%
            \label{fig:NRIEff}
            \includegraphics[width=0.2\textwidth]{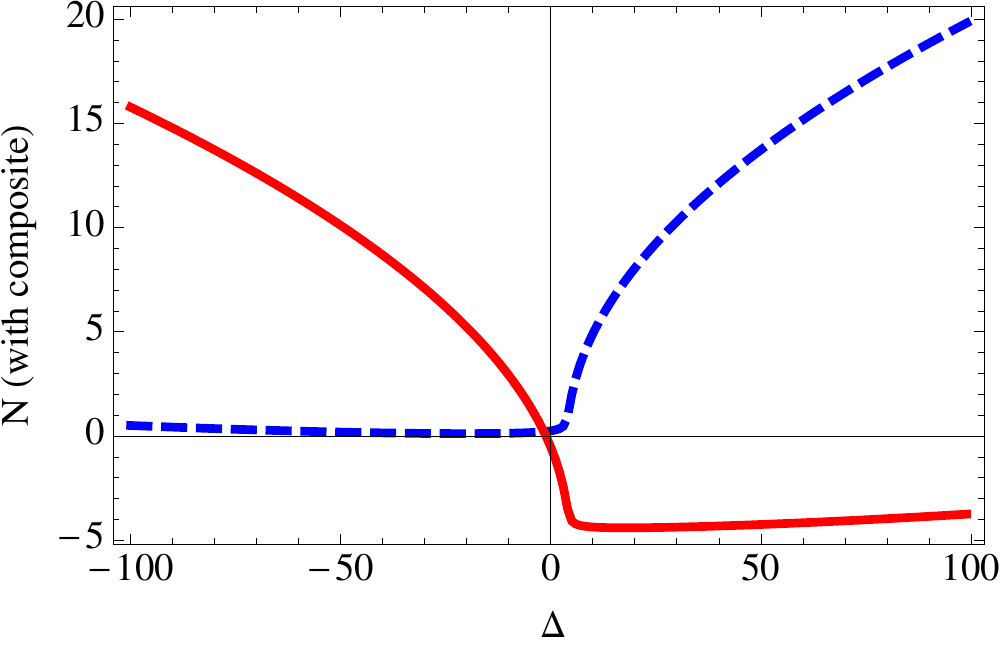}
        }%
               \subfigure[ ]{%
            \label{fig:NRIJurg}
            \includegraphics[width=0.2\textwidth]{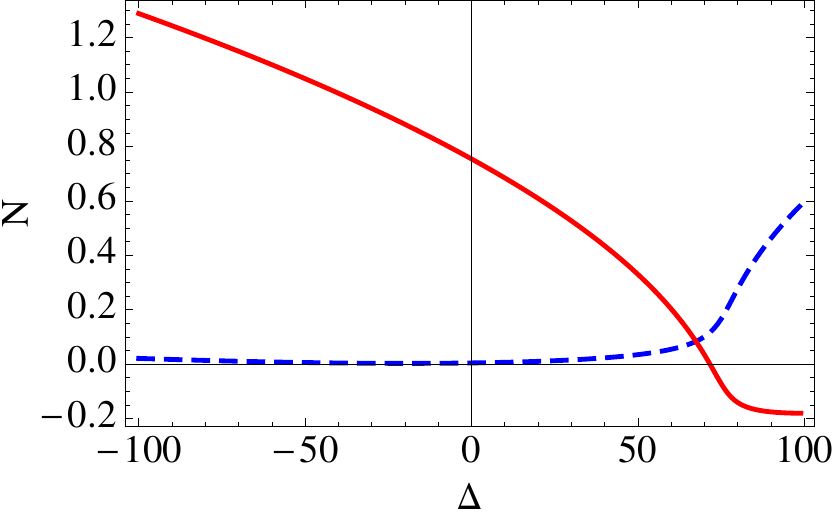}
        }%
    \end{center}
    \caption{%
       Real (red) and imaginary (blue-dashed) refractive indices (a) for a composite system and (b) for a single five level system.  The density of the "negative refractive" system is $\rm{9}\times \rm{10}^{15}\rm{cm}^{-3}$ and the density of the four wave mixing system is $\rm{8}\times \rm{10}^{12}\rm{cm}^{-3}$. $d_1=d_3=8 \times 10^{-2}\rm{cm}$, $d_2=8\rm{cm}$.
     }%
   \label{fig:subfigures454}
\end{figure}

The results of this calculation are given in \fig{fig:454Susceptibilities4WM}.  We can obtain negative refraction using this composite setup. Negative refraction is achieved at lower densities than if we used a single five level system as in \cite{Jurgen}, which is shown in \fig{fig:subfigures454}.  

\begin{figure}[h]
\begin{center}
 {\includegraphics[width=0.45 \textwidth]{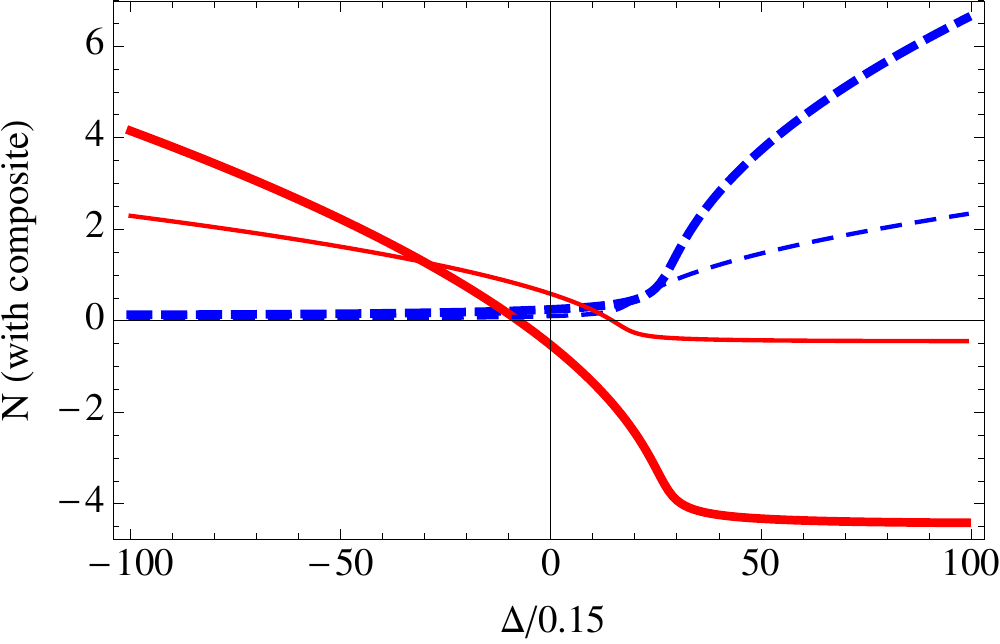}}
\end{center}
\caption{ \protect Refractive index of the system with frequency translation with density of "negative refractive" system $ \rm{9}\times\rm{10}^{15}\rm{cm}^{-3}$ (thick lines) and $ \rm{1}\times\rm{10}^{15}\rm{cm}^{-3}$ (thin lines).  Red - real part of refractive index and blue, dashed - imaginary part of refractive index}
\label{TwoIndices9and5X10^15}
\end{figure}

\subsection{Calculating effective refraction with translation between electric and magnetic fields}

We consider the system given in \fig{Boxesfor545}, with a field of frequency $\omega_1$ which is once again changed to $\omega_2$.  However in this system $\omega_2$ couples an electric dipole transition instead of a magnetic dipole transition, and the results of our calculations are changed.  The setup is given in \fig{Boxesfor545}.

A similar matrix calculation as in section \ref{EffectiveRefractive454} gives us the following effective susceptibilities.  There is cross coupling between fields $E_1$ and $E_2$ in the four level system which is a double $\Lambda$ system.  As EIT effects are required to suppress absorption the population in the ground states of the 4 level system have to be controlled.  This can be accomplished via coupling the ground states to a fifth level.

The effective susceptibilities of this composite system are:

\begin{eqnarray}
\chi_{ee-eff}&=&\frac{c}{\imath \omega_1(d_1 + d_2 + 
  d_3)} (C_3 d_2 d_3 \chi_{22})\nonumber\\
\chi_{eb-eff}&=&\frac{c}{\imath \omega_1(d_1 + d_2 + d_3)}d_2 d_3 C_3 C_6 \chi_{eb}\chi_{21}^{}\nonumber\\
   \chi_{bb-eff}&=&\frac{c}{\imath \omega_1(d_1 + d_2 + d_3)}(C_2 C_6 d_1 d_2 \chi_{be} \chi_{be}(1+\nonumber \\ &&C_4 d_2 \chi_{11}^{})\nonumber\\
\chi_{be-eff}&=&\frac{c}{\imath \omega_1(d_1 + d_2 + d_3)}d_1 d_2 C_2^{} C_4^{}  \chi_{be}^{}\chi_{12}^{}
 \end{eqnarray}
Here $d_1$, $d_2$ and $d_3$ are the thicknesses of the materials in the composite system, $C_2=C_4=C_6= \frac{\imath \omega_2}{c}$ and $C_3= \frac{\imath \omega_1}{c}$.
\subsection{Results}
\begin{figure}[ht!]
     \begin{center}
        \subfigure[ ]{%
            \label{fig:545NN15X10^16ee}
            \includegraphics[width=0.2\textwidth]{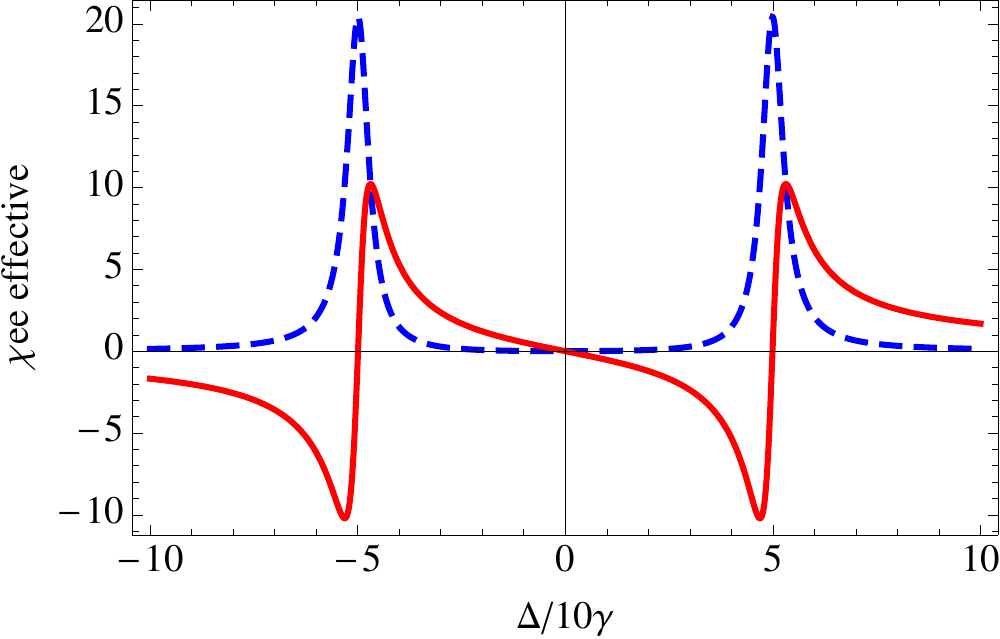}
        }%
        \subfigure[ ]{%
           \label{fig:545NN15X10^15eb}
           \includegraphics[width=0.2\textwidth]{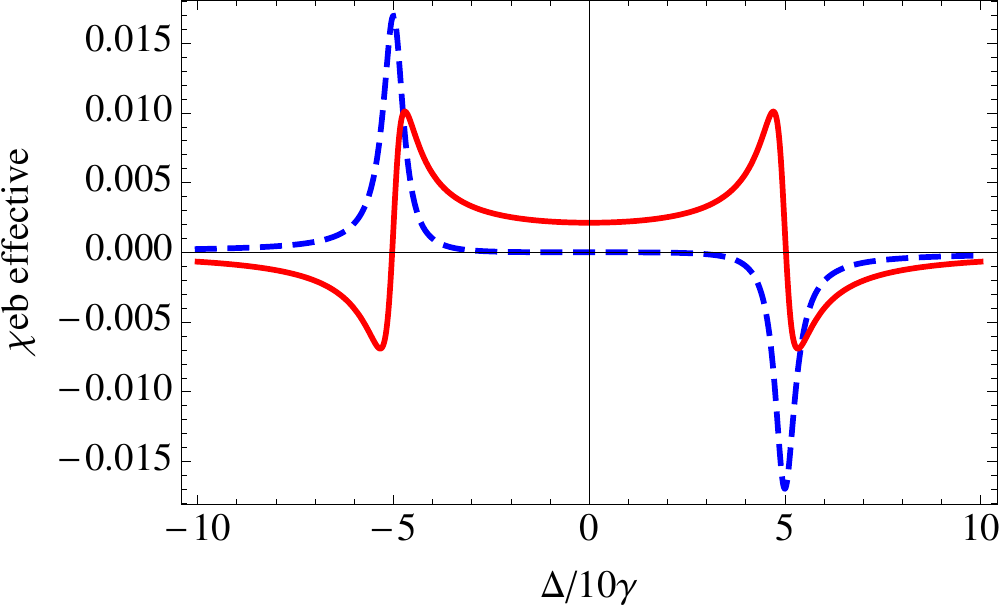}
        } 
        \subfigure[ ]{%
            \label{fig:545NN15X10^15be}
            \includegraphics[width=0.2\textwidth]{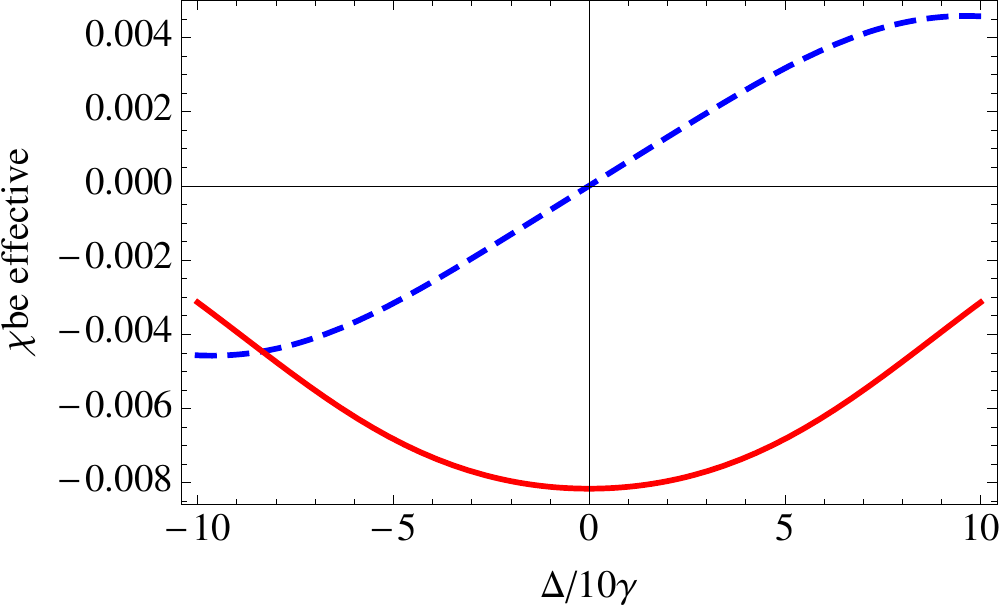}
        }%
        \subfigure[ ]{%
            \label{fig:545NN15X10^15bb}
            \includegraphics[width=0.2\textwidth]{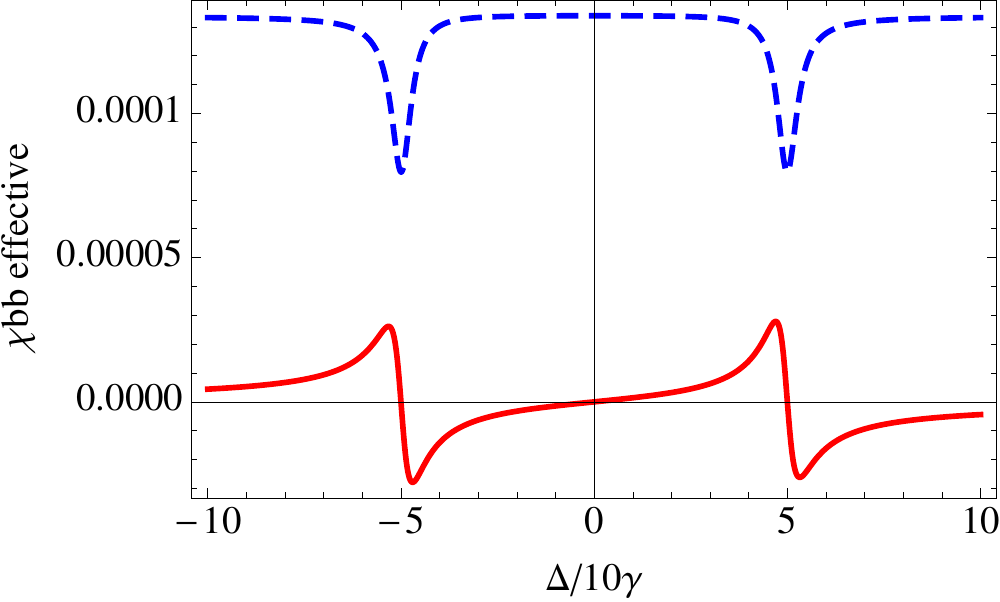}
        }%
    \end{center}
    \caption{%
        (a)-(d)Real(red) and imaginary(blue-dashed) parts of the effective susceptibilities for the magnetic to electric field translational system with density of 4 level system $ \rm{5}\times\rm{10}^{15}\rm{cm}^{-3}$.  The density of the translational system is $ \rm{8}\times\rm{10}^{12}\rm{cm}^{-3}$, and thicknesses are $d_1=d_3=8\times 10^{-2}\rm{cm}$ and $d_2=8\times 10^{-1}\rm{cm}$.
     }%
   \label{fig:545Susceptibilities}
\end{figure}

The results of this calculation are given in \fig{fig:545Susceptibilities}.  Using this system it is possible to obtain enhanced refractive indices with no absorption.  For some detunings it is possible to obtain negative refraction, however in this composite system negative refraction comes at the cost of high absorption.

\begin{figure}[h]
\begin{center}
 {\includegraphics[width=0.45 \textwidth]{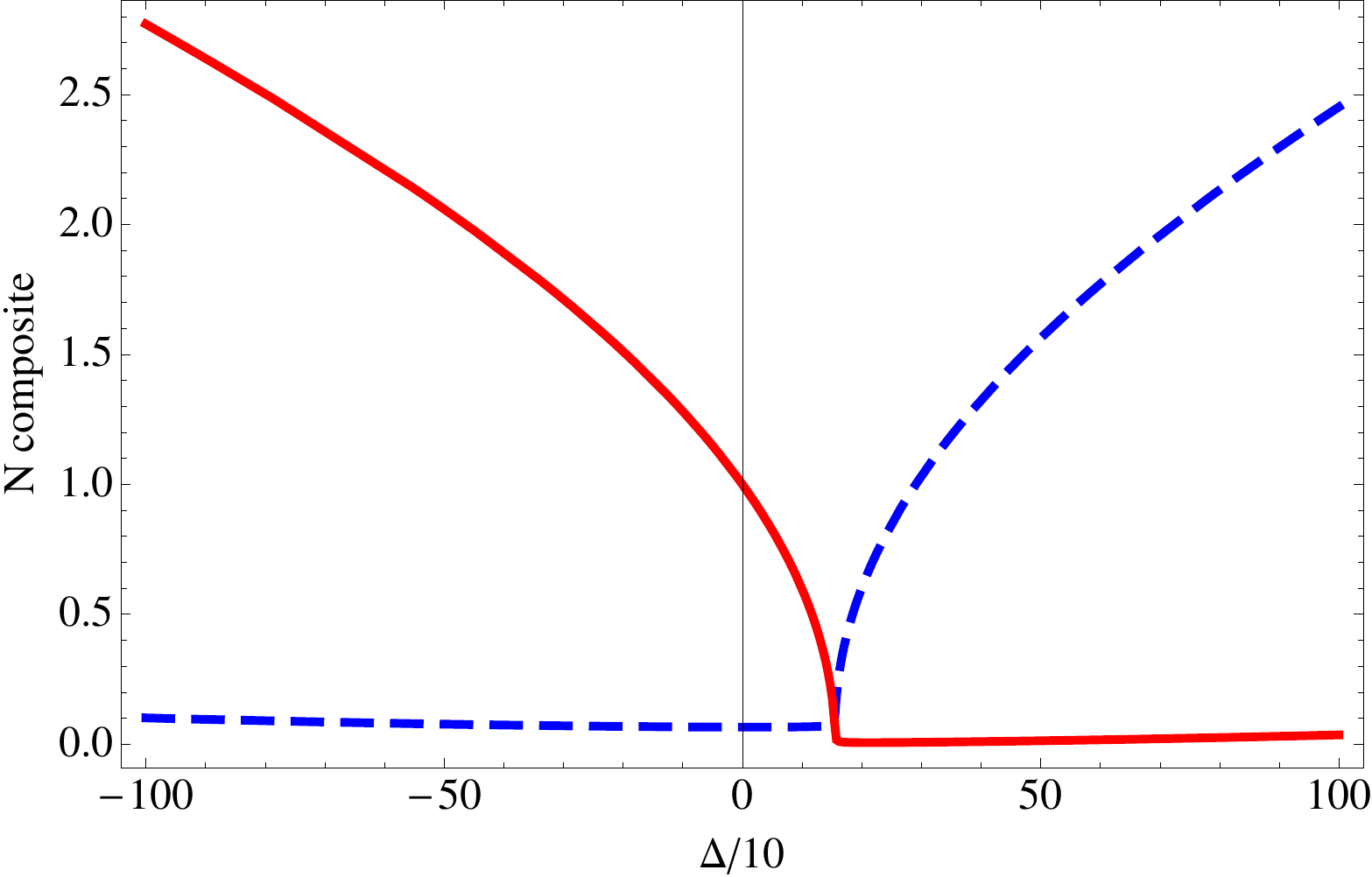}}
\end{center}
\caption{ \protect Refractive index of the system with magnetic to electric field  translation with 4 level scheme density $ \rm{5}\times\rm{10}^{15}\rm{cm}^{-3}$.  Red - real part of refractive index and blue, dashed - imaginary part of refractive index}
\label{545NN5X10^15}
\end{figure}

\section{Conclusion}
The types of materials that can be used for negative refraction has so far been restrained by the necessity of having level structures that have electric and magnetic dipole transitions of the same frequency. It was also necessary to obtain magnetic dipole transitions in the optical regime, which meant that heavy atoms with large level separation needed to be used as magnetic dipole coupling is much weaker compared to electric dipole coupling.

Our investigations show that we can find a combined scheme that generates additional fields in order to obtain the electric and magnetic dipole coupling that is necessary for electromagnetically induced chirality.  We can thus obtain negative refractive index with minimal absorption for materials with a level structures fitting that of the combined system, including atoms, molecules and excitons. As the change of the level structure from a simple 5-level system  to a composite system gives  a great deal of flexibility to the medium we can select, electromagnetically induced chirality can be realized in a wide range of materials.

In conclusion, it can be understood that the scope of negative refractive index materials can be much increased by a  combined level scheme in order to generate additional fields that create electric and magnetic dipole coupling that is necessary for electromagnetically induced chirality. The densities needed to obtain negative refraction are reasonable, and fit the densities of doped crystals and several rare earth material vapors.

\section{Acknowledgements}
We thank M. Fleischhauer and J. Otterbach for useful discussions, and the NSF for funding. 

\section*{Appendix A}
\subsection{Calculating effective refractive Index for a composite system with frequency translation} \label{EffectiveRefractive454}

The incoming electromagnetic field $E_{in}$ creates a field $E_2$, and the exiting fields are given by $E_1$ and $E_2$.  For this system:

\begin{eqnarray}
P_1&=&\chi_{11} E_{1} + \chi_{12} E_{2}\nonumber \\
P_2&=&\chi_{22} E_{2} + \chi_{21} E_{1}\nonumber \\
M_1&=&0 \nonumber \\
M_2&=&0 
\label{box1}
\end{eqnarray}

Here, $P_1= \imath k_1 \frac{d E1}{dz}$ and  $P_2= \imath k_2 \frac{d E2}{dz}$.  There is no magnetic dipole coupling.

If the thickness of the medium of five level system $d_2$,
\begin{equation}
\frac{d E}{dz}=\frac{\imath k_1}{2 \epsilon_0 } P,\label{pol}
\end{equation}
If the length of the medium is assumed to be smaller than one wavelength, the following approximation holds:
\begin{eqnarray}
\frac{\Delta E}{d}&=&\imath \frac{k_1}{2\epsilon_0} P\nonumber \\
\frac{E_{out} -E_{in}}{d}&=&\imath\frac{k_1}{2\epsilon_0} P
\label{approx}
\end{eqnarray}
where P is the polarization density and $d$ is the thickness.  We calculate the magnetization density in a similar manner. 

Using \eq{approx} we can calculate the propagation matrix for the fields:
$\left[\begin{array}{cc}1 &0\\
0&C_2 d_1 \chi_{21}^{} \\
 \end{array}\right]$\\so that
 $\left[\begin{array}{cc}1 & 0\\
0&C_2 d_1 \chi_{21}^{}  \\
 \end{array}\right] \left[\begin{array}{c}H_{in} \\ E_{in} \end{array} \right]=\left[\begin{array}{c}H_{1} \\E_{2} \end{array}\right]$\\\\  Here $C_2=\imath \omega_2 /c$ where $\omega_2$ is the frequency of the created field.
 
 The fields $H_1$ and $E_2$ then enter the second part of the composite system where the polarization and magnetization can be written for the two different frequencies as follows:
 \begin{eqnarray*}
P^{}_A&=&0\\
M^{}_A&=&\chi_{bbA} H^{}_{1} + \chi_{beA} E^{}_{2}\nonumber \\
P^{}_B&=&\chi_{eeA} E^{}_{2} + \chi_{ebA} H^{}_{1}\nonumber \\
M^{}_{B}&=&0\\
\end{eqnarray*}
 
 where A and B are the two frequencies we consider.These equations can be rearranged and written in array form, again using the assumption of small thickness, as: $\left[\begin{array}{c}\frac{H_1^{'} -H_1^{}}{d_2}\\\frac{E_{2}^{} -E_{2}^{}}{d_2} \end{array}\right]=\left[\begin{array}{cc}F_3 \chi_{beA}^{} &F_3 \chi_{bbA}^{}\\ F_4\chi_{eeA}^{}  & F_4 \chi_{ebA}^{} \end{array}\right]\left[\begin{array}{c}H_1^{} \\E_2^{}\end{array}\right]$, which gives us

$\left[\begin{array}{c}H^{'}_{1}\\E^{'}_{2}\end{array}\right]=\left[\begin{array}{cc} F_3 d_2 \chi_{bbA}+1&+F_3 d_2 \chi_{beA}   \\F_4d_2 \chi_{ebA} & F_4 d_2 \chi_{eeA} +1   \end{array}\right]\left[\begin{array}{c}H_{1}\\E_{2}\end{array}\right]$
where $F_3=\imath \omega_1/c $ and $F_4=\imath \omega_2/c$.

The fields then enter a third system which is similar to the first, and the matrix for  the third box can be written as 
$\left[\begin{array}{cc}1  &0\\
0&-C_2 d_3 \chi_{12}\\
 \end{array}\right]$.  By multiplying these matrices together we can find a matrix of the "effective" susceptibilities that transform the field $E_{in}$ ($B_{in}$) to $E_{out}$ ($B_{out}$).

\section*{Appendix B}

\subsection{Maxwell's equations for slowly varying field functions and change of the field phase}

The electromagnetic field radiation when the medium has both polarization and magnetization can be written out at follows

\begin{eqnarray}
\nabla \cdot \mathbf{D}&=&0\\
\nabla \cdot \mathbf{B}&=&0\\
\nabla \times \mathbf{E}&=& - \frac{\partial \mathbf{B}}{\partial t}\\
\nabla \times \mathbf{H}&=& \mathbf{J}+ \frac{\partial \mathbf{D}}{\partial t}
\label{Maxwell}
\end{eqnarray}
where $\mathbf{D}=\epsilon_0 \mathbf{E} + \mathbf{P}$, $\mathbf{B}=\mu_0 \mathbf{H }+\mu_0 \mathbf{M}$ and $\mathbf{J}=\sigma \mathbf{E}$.

We obtain the following results with the aid of Maxwell's equations:
\begin{equation}
-\nabla^2 \mathbf{E}=\mu_0 \sigma \frac{\partial \mathbf{E}}{\partial t} + \mu_0 \epsilon_0 \frac{\partial^2 \mathbf{E}}{\partial t^2}+\mu_0\frac{\partial^2 \mathbf{P}}{\partial t^2}+\mu_0\frac{\partial (\nabla \times \mathbf{M} )}{\partial t}
\label{enabla}
\end{equation}
\begin{equation}
-\nabla^2 \mathbf{B}=\mu_0 \sigma \frac{\partial \mathbf{B}}{\partial t} + \mu_0 \epsilon_0 \frac{\partial^2 \mathbf{B}}{\partial t^2}-\mu_0 \nabla^2 \mathbf{M} +\mu_0\frac{\partial (\nabla \times \mathbf{P} )}{\partial t}
\label{bnabla}
\end{equation}

where $\sigma$ is the conductivity.



The x and y dependence of E, B P and M are neglected and we assume $\mathbf{E}(\mathbf{r},t)=E_x(z,t)\hat{x}+E_y(z,t)\hat{y}$, $\mathbf{B}(\mathbf{r},t)=B_x(z,t)\hat{x}+B_y(z,t)\hat{y}$, $\mathbf{P}(\mathbf{r},t)=P_x(z,t)\hat{x}+ P_y(z,t)\hat{y}$ and $\mathbf{M}(\mathbf{r},t)=M_x(z,t)\hat{x}+ M_y(z,t)\hat{y}$.

Considering equation \eq{enabla}, we then attempt to rewrite the equation separately in each of the vectorial directions.  If we consider the x-direction, \eq{enabla} becomes
\begin{eqnarray}
-\frac{\partial^2 E_x}{\partial z^2} \hat{x}&=&\mu_0 \sigma \frac{\partial E_x}{\partial t}\hat{x} + \mu_0 \epsilon_0 \frac{\partial^2 E_x}{\partial t^2}\hat{x}+\mu_0\frac{\partial^2 P_x}{\partial t^2}\hat{x}\nonumber\\&&+\mu_0\frac{\partial (\partial M_y/\partial z )}{\partial t}\hat{x}
\label{enablax}
\end{eqnarray}

Similarly,\eq{bnabla} becomes

\begin{eqnarray}
-\frac{\partial^2 B_y}{\partial z^2} \hat{y}&=&\mu_0 \sigma \frac{\partial B_y}{\partial t}\hat{y} + \mu_0 \epsilon_0 \frac{\partial^2 B_y}{\partial t^2}\hat{y}-\mu_0\frac{\partial^2 M_y}{\partial z^2}\hat{y}\nonumber\\&&-\mu_0\frac{\partial (\partial P_x/\partial z )}{\partial t}\hat{y}
\label{bnablay}
\end{eqnarray}
in the y direction.  At this point, we remove the directional subscripts and continue the derivations of \eq{enablax} and \eq{bnablay}.


We write the fields of frequency $\nu$ in the following form
\begin{eqnarray}
E_x(z,t)&\equiv&\frac{1}{2}\tilde{E}^+(z,t) e^{-i[\nu t-kz]}+\frac{1}{2}\tilde{E}^-(z,t) e^{+i[\nu t-kz]} \nonumber\\
B_y(z,t)&\equiv&\frac{1}{2}\tilde{B}^+(z,t)e^{-i[\nu t-kz]}+ \frac{1}{2}\tilde{B}^-(z,t)e^{+i[\nu t-kz]}
\label{EBfields}
\end{eqnarray}

We substitute this in \eq{enabla} and \eq{bnabla}where $\phi(z,t)$ is also a slowly varying function of position and time.  The response of the medium, neglecting higher harmonics can be given by the polarization

\begin{eqnarray}
P_x(z,t)&\equiv&\frac{1}{2} \mathcal{P}^+(z,t)e^{-i[\nu t-kz]} + \frac{1}{2} \mathcal{P}^-(z,t)e^{+i[\nu t-kz]} \nonumber\\
\label{PMfields}
\end{eqnarray}
we can derive a similar equation for $M_y(z,t)$.

We make substitutions from \eq{PMfields}, \eq{EBfields} and \eq{EBfields2}, and also use:


$\mathcal{P}(z,t) v^2 \gg \frac{\partial^2 \mathcal{P}(z,t)}{\partial t^2}$ , and which gives us

\begin{equation}
\frac{\partial^2 (\mathcal{P}^+ e^{-i[\nu t-kz]})}{\partial t^2}\approx - \nu^2 \mathcal{P}^+ e^{-i[\nu t-kz]}-2i\nu \frac{\partial\mathcal{P}^+}{\partial t}e^{-i[\nu t-kz]}
\end{equation}


We also make the assumption that $\tilde{E}^+ k^2 \gg \frac{\partial^2 \tilde{E}^+(z,t)}{\partial z^2}$ which gives us the result
\begin{eqnarray}
\frac{\partial^2 \tilde{E}^+(z,t)e^{-i[\nu t-kz]}}{\partial z^2 }&\approx&- k^2\tilde{E}^+(z,t)e^{-i[\nu t-kz]}\nonumber \\
&&+ 2 i k\frac{\partial \tilde{E}^+(z,t)}{\partial z}e^{-i[\nu t-kz]} \nonumber\label{partialEz}
\end{eqnarray}

Once again we make the assumption that $\tilde{E}^+(z,t) v^2 \gg \frac{\partial^2 \tilde{E}^+(z,t)}{\partial t^2}$ which gives us the result
\begin{equation}
\frac{\partial^2 \tilde{E}^+(z,t)}{\partial z^2 }=- v^2 \tilde{E}^+(z,t)e^{-i[\nu t-kz]}+ 2 i v\frac{\partial \tilde{E}(z,t)^+}{\partial z}e^{-i[\nu t-kz]}\nonumber \label{partialEt}
\end{equation}

$ \tilde{E}^+(z,t)$ ( $\tilde{B}^+(z,t)$) is a slowly varying function and can be rewritten in terms of a real amplitude $\mathcal{E}$ ($\mathcal{B}$) and a phase $\phi_E$ ($\phi_B$)

\begin{eqnarray}
\tilde{E}^+(z,t)&=& \mathcal{E}(z,t) e^{i[\phi_E(z,t)]}\nonumber\\
\tilde{B}^+(z,t)&=& \mathcal{B}(z,t) e^{i[\phi_B(z,t)]}
\label{EBfields2}
\end{eqnarray}

The time and space derivatives of $\tilde{E}(z,t)$ can be derived thus:

\begin{equation}
\frac{\partial \tilde{E}^+(z,t)}{\partial z }=i\frac{\partial \phi}{\partial z}\mathcal{E}^+(z,t)e^{i\phi_E (z,t)}+\frac{\partial \mathcal{E}^+(z,t)}{\partial z}e^{i[\phi_E (z,t)]}\nonumber \label{partialEt}
\end{equation}
A similar derivation can be done for $\tilde{B}(z,t)$.

For ease of comprehension, the equations dealing with the electric field will be derived further; the equations of the magnetic fields follow the same derivation.
Substituting \eq{PMfields}and \eq{EBfields} in \eq{enablax}  and adding the necessary subscripts to denote the directions, the resulting equations are

\begin{widetext}
\begin{eqnarray}
k\frac{\partial \mathcal{E}_x} {\partial z}+ \frac{\nu}{c^2} \frac{\partial \mathcal{E}_x}{\partial t}+ k \mathcal{E}_x \frac{\partial \phi_{Ex}} {\partial z}+ \frac{\nu}{c^2} \mathcal{E}_x \frac{\partial \phi_{Ex}}{\partial t} &= &\kappa \mathcal{E}  -\frac{1}{2 \epsilon_0 c^2} \nu^2 \Im (\mathcal{P}_x)- \nonumber \\ && \frac {k \nu}{2\epsilon_0 c^2} \Im (\mathcal{M}_y)+\mathcal{E} (k^2 -\frac{\nu^2}{c^2})  -\nonumber\\&&\frac{1}{2 \epsilon_0 c^2} \nu^2 \Re (\mathcal{P}_x)\nonumber\\&&-\frac{k \nu}{2\epsilon_0 c^2 } \Re (\mathcal{M}_y)
 \label{substitution2}
\end{eqnarray}
\end{widetext}

As $\mathcal{E}$,$\mathcal{B}$,$\mathcal{M}$, $\mathcal{P}$ , $\phi_E$ and $\phi_B$ do not change appreciably in an optical frequency period, we use the slowly varying amplitude and phase approximation and obtain the equations below.
Also, substituting these values in \eq{substitution2} and separating the real and imaginary parts, we obtain the equations below.

\begin{equation}
k\frac{\partial \mathcal{E}_x} {\partial z}+ \frac{\nu}{c^2} \frac{\partial \mathcal{E}_x}{\partial t} =-\kappa \mathcal{E}  -\frac{1}{2 \epsilon_0 c^2 } \nu^2 \Im (\mathcal{P}_x)- \frac {k \nu}{2\epsilon_0 c^2} \Im (\mathcal{M}_y)
\label{field11}
\end{equation}

\begin{equation}
k\frac{\partial \phi_{Ex}} {\partial z}+ \frac{\nu}{c^2} \frac{\partial \phi_{Ex}}{\partial t} =k^2 -\frac{\nu^2}{c^2}  -\frac{1}{2 \epsilon_0 c^2\mathcal{E}_x} \nu^2 \Re (\mathcal{P}_x)-\frac{k \nu}{2\epsilon_0 c^2 \mathcal{E}_x} \Re (\mathcal{M}_y).
\label{phase11}
\end{equation}

A similar derivation gives us the equations for the magnetic field:

\begin{equation}
k \frac{\partial \mathcal{B}_y} {\partial z}+ \frac{\nu}{c^2} \frac{\partial \mathcal{B}_y}{\partial t} =-\kappa \mathcal{B}_y  +\frac{1}{2 \epsilon_0 c^2} k^2 \Im (\mathcal{M}_y)+\frac{k \nu}{2\epsilon_0  c^2} \Im (\mathcal{P}_x).
\label{field22}
\end{equation}

\begin{equation}
k \frac{\partial \phi_{By}} {\partial z}+ \frac{\nu}{c^2} \frac{\partial \phi_{By}}{\partial t} =k ^2-\frac{\nu^2}{c^2} +\frac{1}{2 \epsilon_0 c^2 \mathcal{B}_y} k^2\Re (\mathcal{M}_y)+\frac{k\nu}{2\epsilon_0 c^2 \mathcal{B}_y} \Re (\mathcal{P}_x).
\label{phase22}
\end{equation}
Here $\kappa=\sigma/2 \epsilon_0 c$ is the linear loss coefficient.

These equations can be compared with the equations that are generally used in the study of the interactions of atoms with matter.
\begin{equation}
\frac{\partial \mathcal{E}} {\partial z}+ \frac{1}{c} \frac{\partial \mathcal{E}}{\partial t} =-\kappa \mathcal{E}  -\frac{1}{2 \epsilon_0} k \Im (\mathcal{P})
\label{field11}
\end{equation}

\begin{equation}
\frac{\partial \phi_E} {\partial z}+ \frac{1}{c} \frac{\partial \phi_E}{\partial t} =k -\frac{\nu}{c} -\frac{1}{2 \epsilon_0} k \mathcal{E}^{-1} \Re (\mathcal{P})
\label{phase11}
\end{equation}

and
\begin{equation}
\frac{\partial \mathcal{B}} {\partial z}+ \frac{1}{c} \frac{\partial \mathcal{B}}{\partial t} =-\kappa \mathcal{B}  +\frac{1}{2 \epsilon_0 c^2} k \Im (\mathcal{M})
\label{field21}
\end{equation}

\begin{equation}
\frac{\partial \phi_B} {\partial z}+ \frac{1}{c} \frac{\partial \phi_B}{\partial t} =k -\frac{\nu}{c} +\frac{1}{2 \epsilon_0 c^2} k \mathcal{B}^{-1} \Re (\mathcal{M})
\label{phase21}
\end{equation}

\bibliography{thesisbib1}{}

\begin{thebibliography}{19}
\expandafter\ifx\csname natexlab\endcsname\relax\def\natexlab#1{#1}\fi
\expandafter\ifx\csname bibnamefont\endcsname\relax
  \def\bibnamefont#1{#1}\fi
\expandafter\ifx\csname bibfnamefont\endcsname\relax
  \def\bibfnamefont#1{#1}\fi
\expandafter\ifx\csname citenamefont\endcsname\relax
  \def\citenamefont#1{#1}\fi
\expandafter\ifx\csname url\endcsname\relax
  \def\url#1{\texttt{#1}}\fi
\expandafter\ifx\csname urlprefix\endcsname\relax\def\urlprefix{URL }\fi
\providecommand{\bibinfo}[2]{#2}
\providecommand{\eprint}[2][]{\url{#2}}

\bibitem[{\citenamefont{Kastel et~al.}(2007)\citenamefont{Kastel,
  M.Fleischauer, Yelin, and Walsworth}}]{Jurgen}
\bibinfo{author}{\bibfnamefont{J.}~\bibnamefont{Kastel}},
  \bibinfo{author}{\bibnamefont{M.Fleischauer}},
  \bibinfo{author}{\bibfnamefont{S.}~\bibnamefont{Yelin}}, \bibnamefont{and}
  \bibinfo{author}{\bibfnamefont{R.}~\bibnamefont{Walsworth}},
  \bibinfo{journal}{Phys. Rev. Lett.} \textbf{\bibinfo{volume}{99}},
  \bibinfo{pages}{073602} (\bibinfo{year}{2007}).

\bibitem[{\citenamefont{Veselago}(1968)}]{Veselago}
\bibinfo{author}{\bibfnamefont{V.}~\bibnamefont{Veselago}},
  \bibinfo{journal}{Sov. Phys. Usp} \textbf{\bibinfo{volume}{10}},
  \bibinfo{pages}{509} (\bibinfo{year}{1968}).

\bibitem[{\citenamefont{Pendry et~al.}(1999)\citenamefont{Pendry, Holden,
  Robbins, and Stewart}}]{Pendrytwo}
\bibinfo{author}{\bibfnamefont{J.}~\bibnamefont{Pendry}},
  \bibinfo{author}{\bibfnamefont{A.}~\bibnamefont{Holden}},
  \bibinfo{author}{\bibfnamefont{D.}~\bibnamefont{Robbins}}, \bibnamefont{and}
  \bibinfo{author}{\bibfnamefont{W.}~\bibnamefont{Stewart}},
  \bibinfo{journal}{IEEE Trans. Micro. Theory Tech.}
  \textbf{\bibinfo{volume}{47}}, \bibinfo{pages}{2075} (\bibinfo{year}{1999}).

\bibitem[{\citenamefont{Shelby et~al.}(2001)\citenamefont{Shelby, Smith, and
  Schultz}}]{Shelby}
\bibinfo{author}{\bibfnamefont{R.}~\bibnamefont{Shelby}},
  \bibinfo{author}{\bibfnamefont{D.~R.} \bibnamefont{Smith}}, \bibnamefont{and}
  \bibinfo{author}{\bibfnamefont{S.}~\bibnamefont{Schultz}},
  \bibinfo{journal}{Science} \textbf{\bibinfo{volume}{292}},
  \bibinfo{pages}{77} (\bibinfo{year}{2001}).

\bibitem[{\citenamefont{Shalaev}(2007)}]{Shalaev}
\bibinfo{author}{\bibfnamefont{V.}~\bibnamefont{Shalaev}},
  \bibinfo{journal}{Nature Photonics} \textbf{\bibinfo{volume}{1}},
  \bibinfo{pages}{41} (\bibinfo{year}{2007}).

\bibitem[{\citenamefont{Soukoulis et~al.}(2007)\citenamefont{Soukoulis, Linden,
  and Wegener}}]{Soukoulis}
\bibinfo{author}{\bibfnamefont{C.~M.} \bibnamefont{Soukoulis}},
  \bibinfo{author}{\bibfnamefont{S.}~\bibnamefont{Linden}}, \bibnamefont{and}
  \bibinfo{author}{\bibfnamefont{M.}~\bibnamefont{Wegener}},
  \bibinfo{journal}{Science} \textbf{\bibinfo{volume}{315}},
  \bibinfo{pages}{47} (\bibinfo{year}{2007}).

\bibitem[{\citenamefont{Yen et~al.}(2004)\citenamefont{Yen, Padilla, Fang,
  Vier, Smith, Pendry, Basov, and Zhang}}]{Yen}
\bibinfo{author}{\bibfnamefont{T.~J.} \bibnamefont{Yen}},
  \bibinfo{author}{\bibfnamefont{W.~J.} \bibnamefont{Padilla}},
  \bibinfo{author}{\bibfnamefont{N.}~\bibnamefont{Fang}},
  \bibinfo{author}{\bibfnamefont{D.~C.} \bibnamefont{Vier}},
  \bibinfo{author}{\bibfnamefont{D.~R.} \bibnamefont{Smith}},
  \bibinfo{author}{\bibfnamefont{J.~B.} \bibnamefont{Pendry}},
  \bibinfo{author}{\bibfnamefont{D.~N.} \bibnamefont{Basov}}, \bibnamefont{and}
  \bibinfo{author}{\bibfnamefont{X.}~\bibnamefont{Zhang}},
  \bibinfo{journal}{Science} \textbf{\bibinfo{volume}{303}},
  \bibinfo{pages}{1494} (\bibinfo{year}{2004}).

\bibitem[{\citenamefont{Linden et~al.}(2004)\citenamefont{Linden, Enkrich,
  Wegener, Zhou, Koschny, and Soukoulis}}]{Linden}
\bibinfo{author}{\bibfnamefont{S.}~\bibnamefont{Linden}},
  \bibinfo{author}{\bibfnamefont{C.}~\bibnamefont{Enkrich}},
  \bibinfo{author}{\bibfnamefont{M.}~\bibnamefont{Wegener}},
  \bibinfo{author}{\bibfnamefont{J.}~\bibnamefont{Zhou}},
  \bibinfo{author}{\bibfnamefont{T.}~\bibnamefont{Koschny}}, \bibnamefont{and}
  \bibinfo{author}{\bibfnamefont{C.~M.} \bibnamefont{Soukoulis}},
  \bibinfo{journal}{Science} \textbf{\bibinfo{volume}{303}},
  \bibinfo{pages}{1351} (\bibinfo{year}{2004}).

\bibitem[{\citenamefont{Parimi et~al.}(2004)\citenamefont{Parimi, Lu, Vodo,
  Sokoloff, Derov, and Sridhar}}]{Parimi}
\bibinfo{author}{\bibfnamefont{P.~V.} \bibnamefont{Parimi}},
  \bibinfo{author}{\bibfnamefont{W.~T.} \bibnamefont{Lu}},
  \bibinfo{author}{\bibfnamefont{P.}~\bibnamefont{Vodo}},
  \bibinfo{author}{\bibfnamefont{J.}~\bibnamefont{Sokoloff}},
  \bibinfo{author}{\bibfnamefont{J.~S.} \bibnamefont{Derov}}, \bibnamefont{and}
  \bibinfo{author}{\bibfnamefont{S.}~\bibnamefont{Sridhar}},
  \bibinfo{journal}{Phys. Rev. Lett.} \textbf{\bibinfo{volume}{92}},
  \bibinfo{pages}{127401} (\bibinfo{year}{2004}).

\bibitem[{\citenamefont{Berrier et~al.}(2004)\citenamefont{Berrier, Mulot,
  Swillo, Qiu, Thylen, Talneu, and Anand}}]{Berrier}
\bibinfo{author}{\bibfnamefont{A.}~\bibnamefont{Berrier}},
  \bibinfo{author}{\bibfnamefont{M.}~\bibnamefont{Mulot}},
  \bibinfo{author}{\bibfnamefont{M.}~\bibnamefont{Swillo}},
  \bibinfo{author}{\bibfnamefont{M.}~\bibnamefont{Qiu}},
  \bibinfo{author}{\bibfnamefont{L.}~\bibnamefont{Thylen}},
  \bibinfo{author}{\bibfnamefont{A.}~\bibnamefont{Talneu}}, \bibnamefont{and}
  \bibinfo{author}{\bibfnamefont{A.}~\bibnamefont{Anand}},
  \bibinfo{journal}{Phys. Rev. Lett.} \textbf{\bibinfo{volume}{93}},
  \bibinfo{pages}{073902} (\bibinfo{year}{2004}).

\bibitem[{\citenamefont{Pendry}(2004)}]{Pendry}
\bibinfo{author}{\bibfnamefont{J.}~\bibnamefont{Pendry}},
  \bibinfo{journal}{Science} \textbf{\bibinfo{volume}{306}},
  \bibinfo{pages}{1353} (\bibinfo{year}{2004}).

\bibitem[{\citenamefont{Simmons et~al.}(2012)\citenamefont{Simmons, Proite,
  Miles, Sikes, and Yavuz}}]{YavuzSimmons}
\bibinfo{author}{\bibfnamefont{Z.~J.} \bibnamefont{Simmons}},
  \bibinfo{author}{\bibfnamefont{N.~A.} \bibnamefont{Proite}},
  \bibinfo{author}{\bibfnamefont{J.}~\bibnamefont{Miles}},
  \bibinfo{author}{\bibfnamefont{D.~E.} \bibnamefont{Sikes}}, \bibnamefont{and}
  \bibinfo{author}{\bibfnamefont{D.~D.} \bibnamefont{Yavuz}},
  \bibinfo{journal}{Phys. Rev. B} \textbf{\bibinfo{volume}{85}},
  \bibinfo{pages}{0583810} (\bibinfo{year}{2012}).

\bibitem[{\citenamefont{Kastel et~al.}(2009)\citenamefont{Kastel,
  M.Fleischauer, Yelin, and Walsworth}}]{Jurgen2}
\bibinfo{author}{\bibfnamefont{J.}~\bibnamefont{Kastel}},
  \bibinfo{author}{\bibnamefont{M.Fleischauer}},
  \bibinfo{author}{\bibfnamefont{S.}~\bibnamefont{Yelin}}, \bibnamefont{and}
  \bibinfo{author}{\bibfnamefont{R.}~\bibnamefont{Walsworth}},
  \bibinfo{journal}{Phys. Rev. B} \textbf{\bibinfo{volume}{79}},
  \bibinfo{pages}{063818} (\bibinfo{year}{2009}).

\bibitem[{\citenamefont{O'Dell}(1970)}]{ODell}
\bibinfo{author}{\bibfnamefont{T.}~\bibnamefont{O'Dell}},
  \emph{\bibinfo{title}{The Electrodynamics of Magneto-Electric Media}}
  (\bibinfo{publisher}{Elsevier, NY}, \bibinfo{year}{1970}).

\bibitem[{\citenamefont{McGuinness et~al.}(2010)\citenamefont{McGuinness,
  Raymer, McKinstrie, and Radic}}]{McGuiness}
\bibinfo{author}{\bibfnamefont{H.~J.} \bibnamefont{McGuinness}},
  \bibinfo{author}{\bibfnamefont{M.~G.} \bibnamefont{Raymer}},
  \bibinfo{author}{\bibfnamefont{C.~J.} \bibnamefont{McKinstrie}},
  \bibnamefont{and} \bibinfo{author}{\bibfnamefont{S.}~\bibnamefont{Radic}},
  \bibinfo{journal}{PRL} \textbf{\bibinfo{volume}{105}},
  \bibinfo{pages}{093604} (\bibinfo{year}{2010}).

\bibitem[{\citenamefont{Johnsson and Fleischhauer}(2002)}]{Johnsson2}
\bibinfo{author}{\bibfnamefont{M.}~\bibnamefont{Johnsson}} \bibnamefont{and}
  \bibinfo{author}{\bibfnamefont{M.}~\bibnamefont{Fleischhauer}},
  \bibinfo{journal}{Phys. Rev. B} \textbf{\bibinfo{volume}{66}},
  \bibinfo{pages}{043808} (\bibinfo{year}{2002}).

\bibitem[{\citenamefont{Johnsson and Fleischhauer}(2003)}]{Johnsson}
\bibinfo{author}{\bibfnamefont{M.}~\bibnamefont{Johnsson}} \bibnamefont{and}
  \bibinfo{author}{\bibfnamefont{M.}~\bibnamefont{Fleischhauer}},
  \bibinfo{journal}{Phys. Rev. B} \textbf{\bibinfo{volume}{68}},
  \bibinfo{pages}{023804} (\bibinfo{year}{2003}).

\bibitem[{\citenamefont{Popov et~al.}(2008)\citenamefont{Popov, Myslivets,
  George, and Shalaev}}]{popov1}
\bibinfo{author}{\bibfnamefont{A.~K.} \bibnamefont{Popov}},
  \bibinfo{author}{\bibfnamefont{S.~A.} \bibnamefont{Myslivets}},
  \bibinfo{author}{\bibfnamefont{T.~F.} \bibnamefont{George}},
  \bibnamefont{and} \bibinfo{author}{\bibfnamefont{V.~M.}
  \bibnamefont{Shalaev}}, \bibinfo{journal}{Optics Letters}
  \textbf{\bibinfo{volume}{32}}, \bibinfo{pages}{3044} (\bibinfo{year}{2008}).

\bibitem[{\citenamefont{Lukin et~al.}(1999)\citenamefont{Lukin, Matsko,
  Fleischhauer, and Scully}}]{LukinFourwave}
\bibinfo{author}{\bibfnamefont{M.~D.} \bibnamefont{Lukin}},
  \bibinfo{author}{\bibfnamefont{A.~B.} \bibnamefont{Matsko}},
  \bibinfo{author}{\bibfnamefont{M.}~\bibnamefont{Fleischhauer}},
  \bibnamefont{and} \bibinfo{author}{\bibfnamefont{M.~O.}
  \bibnamefont{Scully}}, \bibinfo{journal}{PRL} \textbf{\bibinfo{volume}{82}},
  \bibinfo{pages}{1847} (\bibinfo{year}{1999}).

\end{thebibliography}
\end{document}